\documentclass[10pt,conference]{IEEEtran}
\makeatletter
\renewcommand{\@IEEEsectpunct}{~}
\makeatother

\usepackage{cite}

\ifCLASSINFOpdf
  \usepackage[pdftex]{graphicx}
\else
  \usepackage[dvips]{graphicx}
\fi
\usepackage{amsmath}
\usepackage{xurl} 
\usepackage{multicol, multirow, tabularx}
\usepackage[table]{xcolor}
\usepackage{url}
\usepackage{csquotes}
\usepackage[normalem]{ulem}
\usepackage{tcolorbox}
\usepackage{caption}
\usepackage[htt]{hyphenat}
\usepackage[hyphenbreaks]{breakurl}
\usepackage{amssymb}
\usepackage[export]{adjustbox} 

\usepackage{url}

\usepackage{csquotes}
\usepackage{threeparttable}
\usepackage{subcaption}
\usepackage{seqsplit}

\begin{document}
%


\title{AssetHarvester: A Static Analysis Tool for Detecting Secret-Asset Pairs in Software Artifacts}





\author{\IEEEauthorblockN{Setu Kumar Basak\IEEEauthorrefmark{1},
K. Virgil English\IEEEauthorrefmark{2}, 
Ken Ogura\IEEEauthorrefmark{3},
Vitesh Kambara\IEEEauthorrefmark{4},
Bradley Reaves\IEEEauthorrefmark{5} and
Laurie Williams\IEEEauthorrefmark{6}}
\IEEEauthorblockA{North Carolina State University, USA\\
\IEEEauthorrefmark{1}sbasak4@ncsu.edu,
\IEEEauthorrefmark{2}kvenglis@ncsu.edu,
\IEEEauthorrefmark{3}kogura@ncsu.edu,
\IEEEauthorrefmark{4}vkkambar@ncsu.edu,
\IEEEauthorrefmark{5}bgreaves@ncsu.edu,
\IEEEauthorrefmark{6}lawilli3@ncsu.edu}}

\maketitle

\begin{abstract}
GitGuardian monitored secrets exposure in public GitHub repositories and reported that developers leaked over 12 million secrets (database and other credentials) in 2023, indicating a 113\% surge from 2021. Despite the availability of secret detection tools, developers ignore the tools' reported warnings because of false positives (25\%-99\%). However, each secret protects assets of different values accessible through asset identifiers (a DNS name and a public or private IP address). The asset information for a secret can aid developers in filtering false positives and prioritizing secret removal from the source code. However, existing secret detection tools do not provide the asset information, thus presenting difficulty to developers in filtering secrets only by looking at the secret value or finding the assets manually for each reported secret. \textit{The goal of our study is to aid software practitioners in prioritizing secrets removal by providing the assets information protected by the secrets through our novel static analysis tool.} We present AssetHarvester, a static analysis tool to detect secret-asset pairs in a repository. Since the location of the asset can be distant from where the secret is defined, we investigated secret-asset co-location patterns and found four patterns. To identify the secret-asset pairs of the four patterns, we utilized three approaches (pattern matching, data flow analysis, and fast-approximation heuristics). We curated a benchmark of 1,791 secret-asset pairs of four database types extracted from 188 public GitHub repositories to evaluate the performance of AssetHarvester. AssetHarvester demonstrates precision of (97\%), recall (90\%), and F1-score (94\%) in detecting secret-asset pairs. Our findings indicate that data flow analysis employed in AssetHarvester detects secret-asset pairs with 0\% false positives and aids in improving the recall of secret detection tools. Additionally, AssetHarvester shows 43\% increase in precision for database secret detection compared to existing detection tools through the detection of assets, thus reducing developer's alert fatigue.

\end{abstract}


%
\IEEEpeerreviewmaketitle

\section{Introduction} \label{Introduction}
\begin{figure}
    \centering
    \begin{subfigure}[b]{\columnwidth}
        \centering
        \includegraphics[width=\columnwidth, frame]{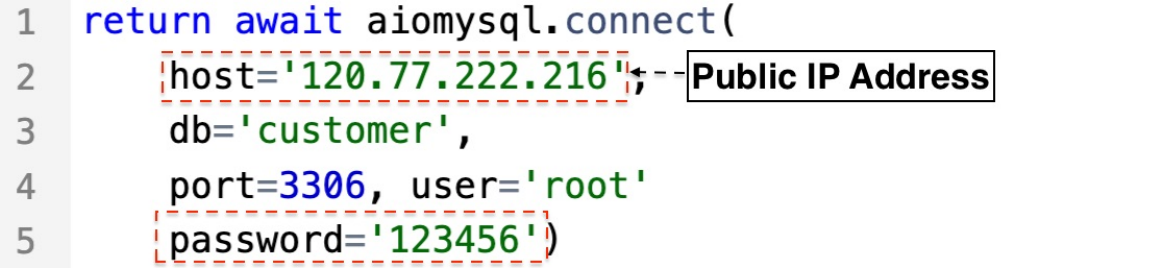}
        \caption[]{{A public IP address protected by a secret}}
        \label{fig:public-ip-address}
    \end{subfigure}
\vskip\baselineskip
    \begin{subfigure}[b]{\columnwidth}
        \centering
        \includegraphics[width=\columnwidth, frame]{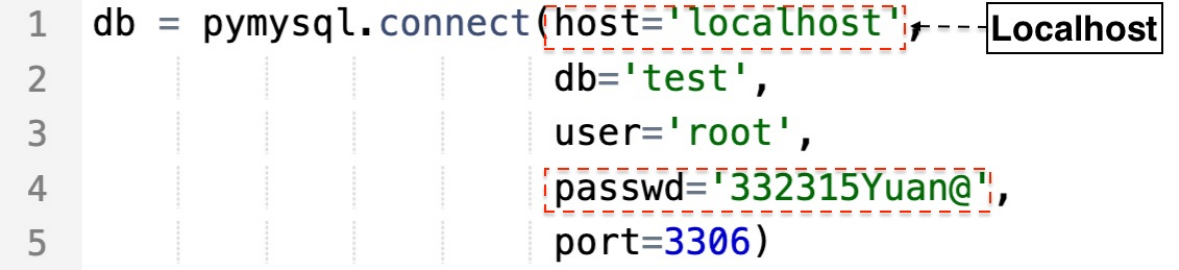}
        \caption[]{{A localhost protected by a secret}}
        \label{fig:localhost}
    \end{subfigure}
\caption{A secret can protect both real and non-sensitive assets, such as a public IP address (real) and localhost (non-sensitive).}
\label{fig:asset-examples-intro}
\end{figure}

In March 2024, GitGuardian reported a 113\% surge in developers leaking over 12 million secrets in public GitHub repositories in 2023 compared to 2021~\cite{gitguardian-secret-sprawl}. They found that 1.7 million authors leaked secrets out of 14.9 million who pushed code to GitHub in 2023. Secrets, such as database credentials and API keys, are essential for integrating with external services such as customer databases and payment systems. However, developers keep hard-coded secrets in application packages and version control systems (VCS), exposing sensitive information to attackers~\cite{meli2019bad,android-leak}. For example, an attacker leveraged hard-coded credentials present in Uber's PowerShell script and launched an account takeover of their internal tools and productivity applications in September 2022~\cite{uber-breach}.

At present, many open-source and proprietary secret detection tools, such as TruffleHog~\cite{trufflehog} and GGShield~\cite{ggshield}, are available to prevent leaking secrets. However, Basak et al.~\cite{basak2023compare} investigated five open-source and four proprietary secret detection tools and found that five of these nine tools demonstrate a precision of less than 7\%. The tool with the highest precision (75\%) among the nine tools misses many secrets, having only 3\% recall. Thus, developers may develop ``alert-fatigue"~\cite{alert-fatigue} and start to ignore the warnings reported by the tools. 

A secret in a software artifact protects an asset (a database or an API service) accessible through asset identifiers (a URL, a DNS name, or an IP address). However, a secret may look like a false positive that protects a real asset. For example, Figure~\ref{fig:public-ip-address} shows a customer database with a public IP address (``120.77.222.216") protected by the password ``123456". On the contrary, a secret may look like a true positive but protects a non-sensitive asset. For example, Figure~\ref{fig:localhost} shows the password ``332315Yuan@" protects a test database of a ``localhost" that is typically not vulnerable to outside attackers. However, existing secret detection tools do not provide the asset information corresponding to a secret. As a result, developers manually filter alerts based on the secret value without the asset information and may ignore a secret protecting a valuable asset. In addition, developers may lose their development time while manually identifying the asset for each secret reported by the tools. Thus, programmatically identifying the assets protected by the secrets can aid in reducing the manual effort of developers to filter false positives and identify secrets that protect valuable assets. Additionally, developers can prioritize efforts to remove secrets based on the asset context.

\textit{The goal of our study is to aid software practitioners in prioritizing secrets removal by providing the assets information protected by the secrets through our novel static analysis tool.}

However, identifying the assets protected by the secrets is not straightforward since the asset identifiers can have multiple parts defined separately in the source code. For example, a database server address contains a host, port, and database name. In addition, multiple assets can be present in the same file (such as a configuration file). Additionally, the asset can be distant from where the secret is defined in the source code. For example, a secret is disclosed in one file, and the corresponding asset is disclosed in another file of the repository. Thus, even if we identify the asset, mapping the asset to the correct secret is challenging. In this study, we investigate how we can programmatically identify both the secret and the asset protected by the secret and provide answers to our research questions:

\begin{itemize}
  \item \textbf{RQ1:} What are the secret-asset co-location patterns present in software artifacts? (Section~\ref{AssetPatterns})
  \item \textbf{RQ2:} What performance can be achieved in detecting assets protected by secrets via static analysis in terms of precision, recall, and F1-score? (Section~\ref{Results})
\end{itemize}

We curated AssetBench, a benchmark of 1,791 secret-asset pairs of four database types extracted from 188 public GitHub repositories. To answer RQ1, we investigated and categorized the secret-asset co-location patterns in the source code. To answer RQ2, we constructed AssetHarvester, a static analysis tool to identify the database secret-asset pairs. In constructing AssetHarvester, we utilized pattern matching, data flow analysis, and fast-approximation heuristics to detect the secret-asset pairs. We evaluated the performance of AssetHarvester against AssetBench in terms of precision, recall, and F1-score. We provide a summary of our contributions as follows:

\begin{itemize}
    \item We constructed AssetHarvester, a static analysis tool to detect the assets protected by secrets to aid developers in prioritizing secrets removal. Additionally, AssetHarvester has shown 43\% and 50\% increase in precision and recall, respectively, for database secret detection compared to existing detection tools through the detection of assets.
    \item We have made the implementation of AssetHarvester publicly available~\cite{artifacts}. Additionally, we provided AssetBench, a dataset of secret-asset pairs that can be extended and utilized by researchers and tool developers for future research and tool development.
\end{itemize}

The rest of our paper is structured as follows: Section~\ref{AssetTypeSelection} introduces the selection process of asset types, followed by the benchmark dataset curation and the secret-asset co-location patterns. We discuss the AssetHarvester construction and evaluation results of AssetHarvester against AssetBench in Sections~\ref{AssetHarvester} and~\ref{Results}, followed by the implications of our work. We discuss the ethics and limitations of our study in Sections~\ref{Ethics} and~\ref{ThreatToValidity}, respectively. We discuss the related work in Section~\ref{RelatedWork} and conclude in Section~\ref{Conclusion}. 

\section{Asset Type Selection} \label{AssetTypeSelection}
\begin{figure}[]
    \includegraphics[width=\columnwidth, frame]{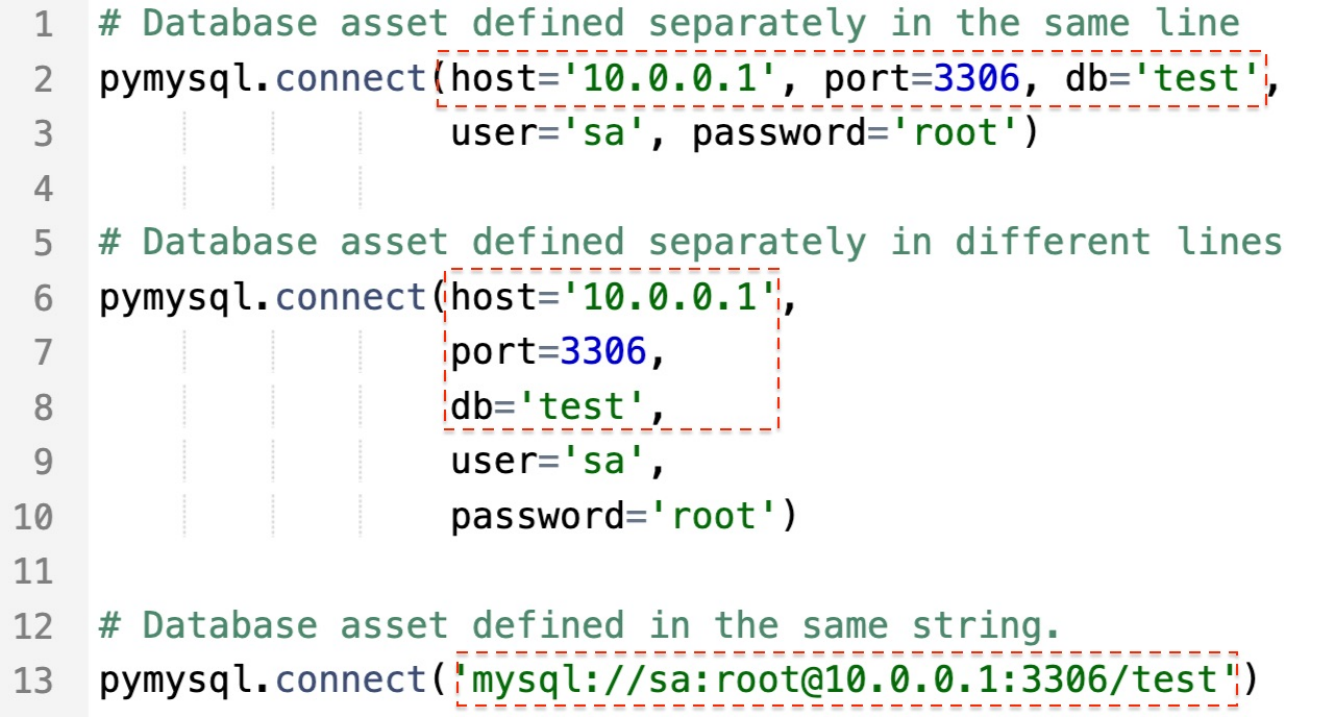}
    \caption{The database asset identifier has three parts (host, port, and database name) that are defined separately in the same line, in separate lines, or together in the same string.}
    \label{fig:database-asset-structure}  
\end{figure}

A software artifact may contain different types of assets, such as database server addresses and API URLs, which are protected by secrets. However, the 2024 GitGuardian report~\cite{gitguardian-secret-sprawl} reveals that out of 12 million exposed secrets in GitHub, the top secret type is database providers. Additionally, database assets can be challenging to detect due to multiple asset identifier formats, among other asset types. For example, Figure~\ref{fig:database-asset-structure} shows a database asset identifier can have multiple parts (host, port, and database name) defined separately in the same line (line 2) or different lines (lines 6-8). The database asset can also be in the same string (line 13). Thus, we selected database secret-asset pairs to be detected in this study. Since multiple database providers are present, we need to narrow our scope to maintain our study's feasibility. We observed that the top five databases developers use are PostgreSQL~\cite{postgresql}, MySQL~\cite{mysql}, SQLite~\cite{sqlite}, MongoDB~\cite{mongodb}, and SQL Server~\cite{sqlserver}, according to the Stack Overflow Developer Survey 2023~\cite{SOdevelopersurvey2023}. However, we excluded SQLite from our study since SQLite is a file-based database requiring no authentication. Finally, we selected these four databases for our study.

\section{AssetBench} \label{BenchmarkDataset}
To create a dataset of secret-asset pairs, we started with SecretBench~\cite{secretbench}, a publicly available benchmark dataset of software secrets. We accessed the dataset through Google Cloud Storage (Bucket Name: \textit{secretbench}) and Google BigQuery (Dataset ID: \textit{\seqsplit{dev-range-332204.secretbench.secrets}}). The authors curated 818 repositories from the September 2022 snapshot of Google BigQuery Public Dataset of GitHub~\cite{google-big-query} (Dataset ID: \textit{bigquery-public-data.github\_repos}). The repositories in the dataset comprise source codes of 49 programming languages. The secrets present in the repositories are extracted using two open-source secret detection tools, TruffleHog~\cite{trufflehog} and Gitleaks~\cite{gitleaks}. The dataset contains 97,479 labeled plain-text secrets, manually labeled as true or false by two authors of the SecretBench. In addition, the dataset provides additional metadata such as repository name, commit ID, file path, and line number where the secrets have been found. However, the dataset does not provide information regarding the assets protected by the secrets. Hence, we extended the dataset as \textit{AssetBench} by identifying assets for each secret in our study. 

\textbf{Filtering Dataset:} Before identifying assets, we applied the following selection criteria to filter SecretBench.

\textit{\uline{Criteria 1 (Programming Language):}} The 2023 GitGuardian report indicates that developers most frequently leaked secrets in repositories written in Python programming language~\cite{gitguardian-secret-sprawl-2023}. Additionally, we observed that Python is third among 49 programming languages containing secrets in SecretBench. Thus, we chose repositories containing Python source code for our study. We selected 188 repositories from 818 repositories and 34,569 secrets from 97,479 secrets of SecretBench.     

\textit{\uline{Criteria 2 (Database Type):}} A repository can contain different types of secrets, such as API keys and database credentials. We filtered the secrets of the selected four databases (Section~\ref{AssetTypeSelection}) and selected 2,114 secrets from 34,569 secrets. 

\textbf{Identifying Assets:} Next, the first and third authors manually inspected each secret independently using the repository name, commit ID, file path, and line number provided by the dataset and identified the asset protected by the secret. However, the asset may not be present in the same file where the secret is located. In such cases, both authors inspected the candidate asset-containing files in the repository. The asset's value for each secret with additional metadata (the file path and line number where the asset is found) is collected. We observed the agreement of finding the secret-asset pairs with a Cohen's Kappa~\cite{cohen-kappa} score of 0.96 between two authors, which indicates a ``near perfect agreement" according to Landis and Koch's interpretation~\cite{landis-koch}. The disagreements were resolved after a discussion between the two authors. However, neither author found corresponding assets for 323 secrets. We removed those secrets and selected 1,791 secret-asset pairs. In Table~\ref{asset-types}, we presented the database type and the number of secret-asset pairs with percentages for each type. AssetBench contains 25 secret-asset pairs for SQL Server, representing 1.4\% of the total pairs. The relatively lower percentage might stem from SQL Server's proprietary nature, leading to lesser adoption in open-source projects than available open-source databases.

\begin{table}
\begin{center}
\caption{Count of Secret-Asset pairs for four databases}
\label{asset-types}
\small
\begin{tabular}{|c| c| c|} 
 \hline
 \textbf{Database Type} & \textbf{\# Secret-Asset Pair} & \textbf{\% of Pair} \\ [0.5ex] 
 \hline\hline
 MySQL & 777 & 43.4\%\\ 
 \hline
 PostgreSQL & 679 & 37.9\%\\
 \hline
 MongoDB & 310 & 17.3\%\\
 \hline
 SQL Server & 25 & 1.4\%\\ 
 \hline
\end{tabular}
\end{center}
\end{table}

\textbf{Developer Survey:} To evaluate whether the committer of the secret agrees with our identified asset for the secret, we conducted a developer survey. First, to avoid recall bias~\cite{recall-bias}, we selected secret-asset pairs committed between 2021 and 2022 and identified 683 secret-asset pairs. Next, we filtered out secret-asset pairs that have a noreply (\texttt{xxx@users.noreply.github.com}) or GitHub Actions bot (\texttt{action@github.com}) commit email address~\cite{github-no-reply} and selected 490 secret-asset pairs. Next, we randomly selected 100 secret-asset pairs to avoid selection bias~\cite{selection-bias} and emailed the developers to know their agreement with our labeling and the reason for any disagreements. In the email, we provided the secret-asset pair information with a screenshot of the code where the secret-asset pair is found. We received 27 responses out of 100, and all respondents agreed with our label.

\textbf{Dataset Storage:} Our curated dataset, AssetBench, is stored in Google BigQuery (Dataset ID: \textit{\seqsplit{dev-range-332204.assetbench.assets}}) as a relational structured data. Users can access the dataset through SQL queries. However, due to the sensitive nature of the dataset, we will provide access to the dataset to only selected researchers and tool developers. Those who require access need to contact us through email.

\section{Secret-Asset Co-location Patterns} \label{AssetPatterns}
\begin{figure*}[h]
    \centering
    \begin{subfigure}[b]{0.49\textwidth}
        \centering
        \includegraphics[width=\textwidth, frame]{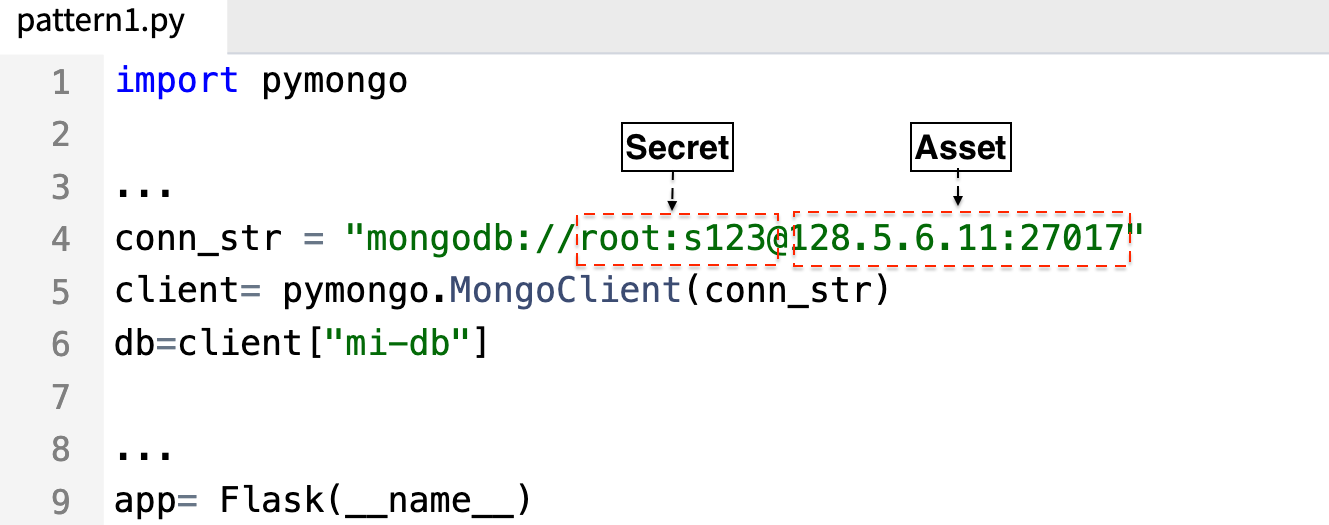}
        \caption[]{{Pattern 1 (Same String, Same Line, Same File)}}
        \label{fig:pattern1}
    \end{subfigure}
\hfill
    \begin{subfigure}[b]{0.49\textwidth}
        \centering
        \includegraphics[width=\textwidth, frame]{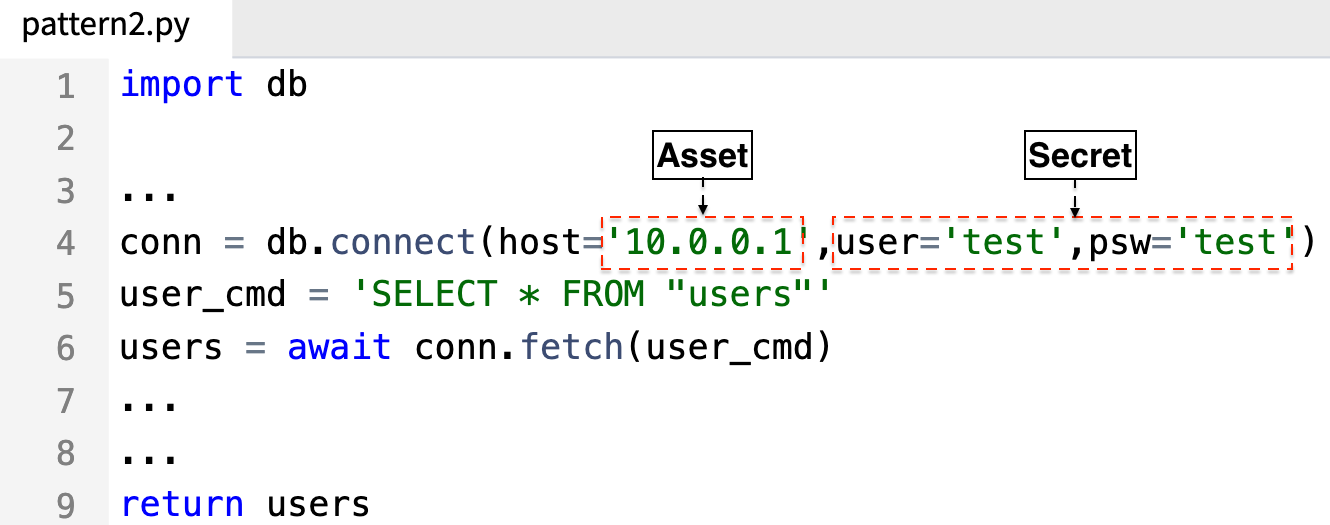}
        \caption[]{{Pattern 2 (Separate Strings, Same Line, Same File)}}
        \label{fig:pattern2}
    \end{subfigure}
\vskip\baselineskip
    \begin{subfigure}[b]{0.49\textwidth}
        \centering
        \includegraphics[width=\textwidth, frame]{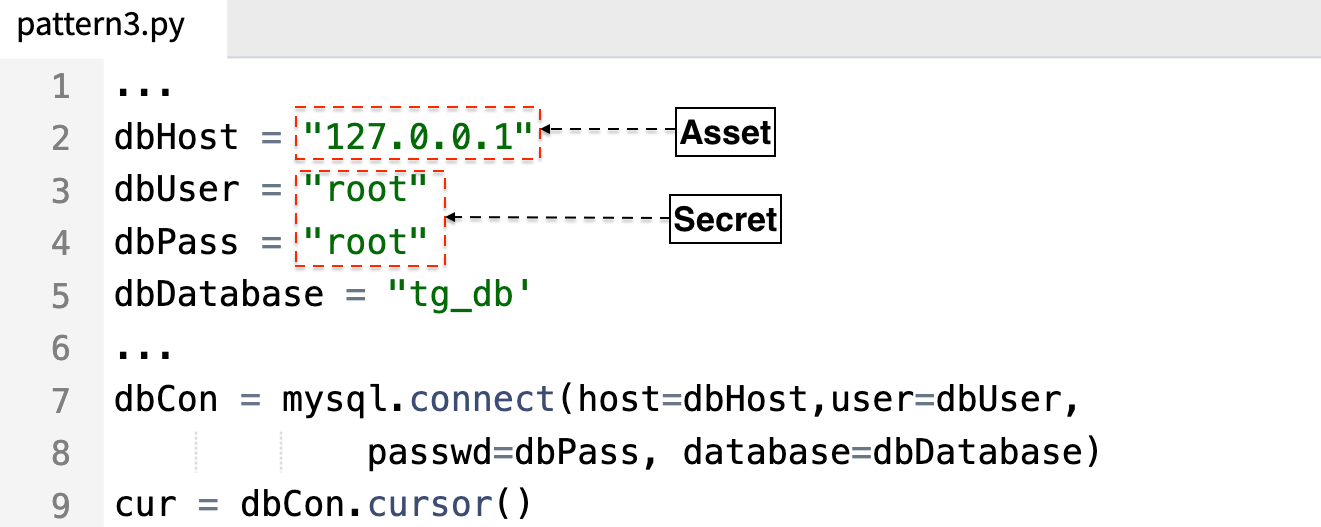}
        \caption[]{{Pattern 3 (Separate Strings, Separate Lines, Same File)}}
        \label{fig:pattern3}
    \end{subfigure}
\hfill
    \begin{subfigure}[b]{0.49\textwidth}
        \centering
        \includegraphics[width=\textwidth, frame]{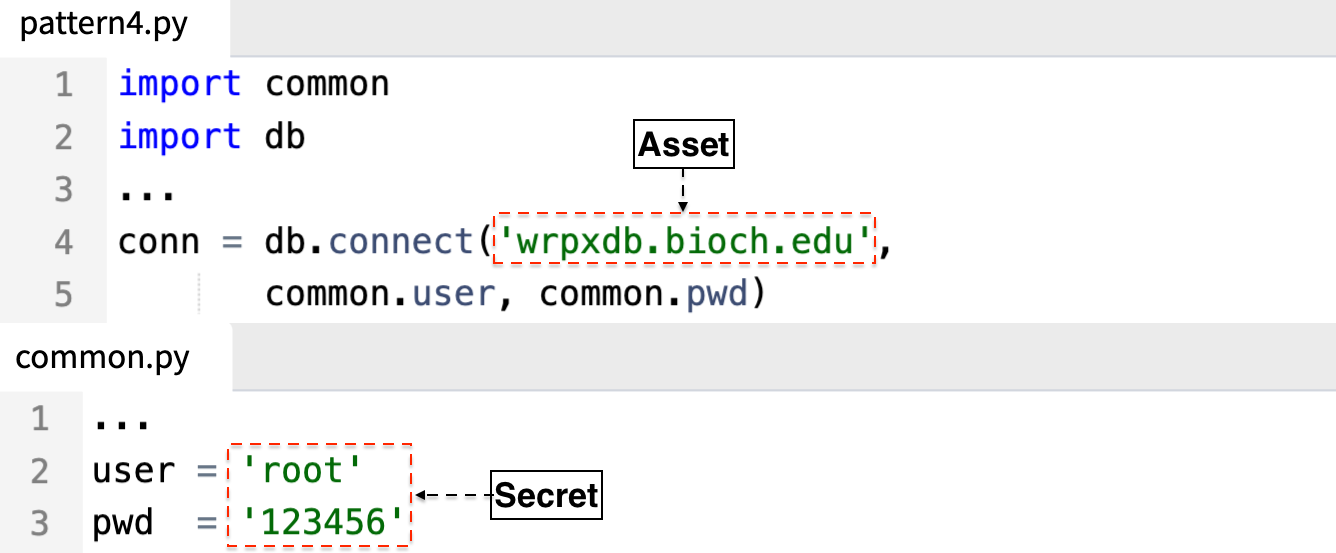}
        \caption[]{{Pattern 4 (Separate Strings, Separate Lines, Separate Files)}}
        \label{fig:pattern4}
    \end{subfigure}
\caption{We identified four types of secret-asset co-location patterns in the source code.}
\label{fig:aenb}
\end{figure*}

To answer RQ1, the first and second authors independently inspected a random sample of 100 secret-asset pairs from the AssetBench. Both authors observed the location pattern of a secret and the corresponding asset and identified four mutually exclusive secret-asset co-location patterns. We utilized the co-location patterns for programmatically identifying secret-asset pairs in the construction of AssetHarvester, as described in Section~\ref{AssetHarvester}. We now describe the four secret-asset co-location patterns. The number in parenthesis denotes the occurrences of each pattern found out of 100 secret-asset pairs.

\textbf{Pattern 1 (Same String, Same Line, Same File) (54):} The secret and the corresponding asset identifier can be present in the same string and the same line of a file, such as in a database connection string. For example, Figure~\ref{fig:pattern1} shows a MongoDB connection string (\texttt{\seqsplit{"mongodb://root:s123@128.5.6.11:27017"}}) defined in line 4, where \texttt{"root"} and \texttt{"s123"} are the username and password of the database, and \texttt{\seqsplit{"128.5.6.11:27017"}} is the database server address.

\textbf{Pattern 2 (Separate Strings, Same Line, Same File) (20):} The secret and the corresponding asset identifier can be present in the same line of the file but defined separately in multiple strings. For example, the database server address (\texttt{\seqsplit{"10.0.0.1"}}), the username (\texttt{"test"}), and the password (\texttt{"test"}) are defined and passed as separate arguments to \texttt{db.connect} method in line 4, as shown in Figure~\ref{fig:pattern2}.

\textbf{Pattern 3 (Separate Strings, Separate Lines, Same File) (19):} The secret and the corresponding asset identifier can be present in the same file of the repository but defined in separate strings and separate lines. For example, as shown in Figure~\ref{fig:pattern3}, the username (\texttt{"root"}) and password (\texttt{"root"}) are defined in lines 3 and 4, respectively, whereas the database server address (\texttt{"127.0.0.1"}) is defined in line 2 of the same file.

\textbf{Pattern 4 (Separate Strings, Separate Lines, Separate Files) (7):} The secret and the corresponding asset identifier can be present in separate files of the repository. For example, as shown in Figure~\ref{fig:pattern4}, the username (\texttt{"root"}) and password (\texttt{"123456"}) of the database are defined in lines 2 and 3 of \texttt{common.py} file, respectively. The values of the \texttt{common.py} file are imported in line 1 of the \texttt{pattern4.py} file, where the database server address (\texttt{\seqsplit{"wrpxdb.bioch.edu"}}) is defined in line 4. However, the secret and the asset may not always be present in the same file types. For example, both the files in Figure~\ref{fig:pattern4} have \texttt{.py} extension. However, the secret or asset can be defined in configuration files such as \texttt{config.yml} file, that can be read in a \texttt{.py} file.

\section{AssetHarvester} \label{AssetHarvester}
We utilized the identified secret-asset co-location patterns (Section~\ref{AssetPatterns}) and constructed AssetHarvester using pattern matching (Step 1), data flow analysis (Step 2), and fast-approximation heuristic (Step 3). We now discuss the three-step process of constructing AssetHarvester.

\textbf{Step 1: Asset Finding Using Pattern Matching:}\label{PatternMatching} We observed from Pattern 1 in Section~\ref{AssetPatterns} that the secret and the corresponding database asset are present in a database connection string. Since a database connection string follows a specific format, we can formulate regular expressions (regex) to identify the secret and the corresponding asset. We now discuss our approach to formulating the regex and identifying the assets protected by the corresponding secrets.

\textbf{Step 1.1 Formulating Regex:} In this step, we manually inspected the documentation for each database type and identified the database connection string format. Then, we categorized the connection string formats into three groups and formulated the regex for each group, as shown in Table~\ref{regex-database}. We now discuss how we categorized the database connection string formats and formulated regex for each group.

\begin{table*} [!htb]
\footnotesize
\centering
\caption{List of regexes categorized into three groups for identifying database connection string}
\label{regex-database}
    \includegraphics[width=\textwidth]{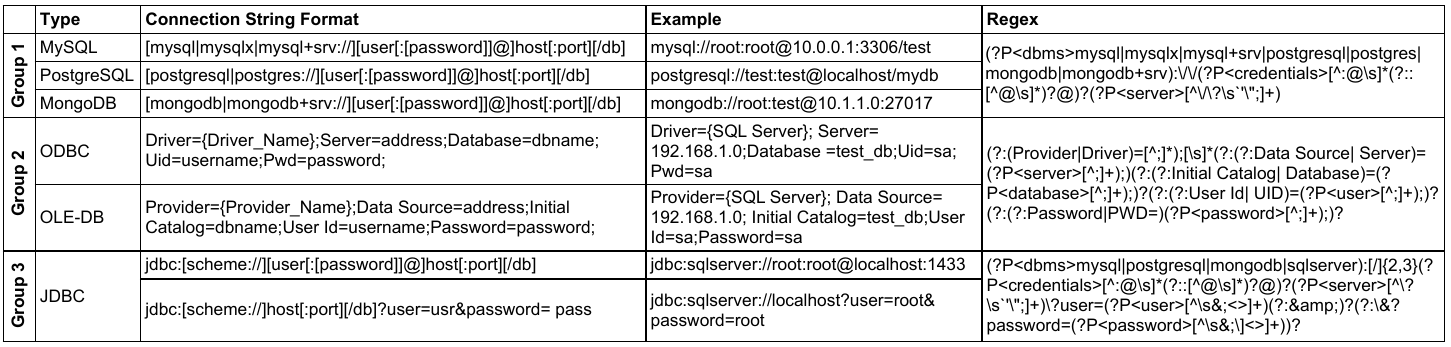}
\end{table*}

\textit{\uline{Group 1: (MySQL, PostgreSQL \& MongoDB):}} We observed that according to MySQL~\cite{mysqlconnstr}, PostgreSQL~\cite{postgresqlconnstr}, and MongoDB~\cite{mongodbconnstr} documentations, these three database types have similar connection string formats. The common format is \textit{``[scheme://][user[:[password]]@]host[:port][/db]"}. The \texttt{scheme} refers to the transport protocol, \texttt{user:password} refers to the credentials, \texttt{host:port} refers to the server address, and \texttt{db} is the database name in the connection. We observed that only the scheme type differs in the three database connection strings. For example, MySQL uses \texttt{"mysql"} or \texttt{"mysqlx"} whereas PostgreSQL uses \texttt{"postgresql"} or \texttt{"postgres"} as the scheme types. Hence, we formulated a common regex with the different scheme types for Group 1.

\textit{\uline{Group 2: (ODBC \& OLE-DB):}} The Open Database Connectivity (ODBC)~\cite{odbc} and the Object Linking and Embedding Database (OLE-DB)~\cite{oledb} are two standard APIs that provide support for accessing and interacting with different databases. ODBC and OLE-DB support the selected four databases in our study. We noticed that ODBC and OLE-DB have similar connection string formats consisting of key-value pairs separated by semicolons. For example, the connection string format for ODBC is \textit{``Driver={Driver\_Name}; Server=address; Database=dbname; Uid=username; Pwd=password;"}. The key \texttt{Driver} refers to the ODBC database driver to be used such as ``SQL Server". The key \texttt{Server} refers to the address of the database server, and keys \texttt{Uid} and \texttt{Pwd} refer to the credentials in the connection. However, for OLE-DB, the key for the database server is \texttt{Data Source} instead of \texttt{Server}. Hence, we formulated a common regex with all the ODBC and OLE-DB key-value pairs for Group 2.

\textit{\uline{Group 3: (JDBC):}} Similar to ODBC and OLE-DB, Java Database Connectivity (JDBC)~\cite{jdbc} is a standard API that allows Java applications to interact with different databases. In our study, though we selected repositories containing Python programming language, repositories can have Java source code containing JDBC connection strings. In addition, Packages such as \texttt{JayDeBeApi}~\cite{jaydeapi} are available that allow the connection of a database using the JDBC connection string from the Python source code. As shown in Table~\ref{regex-database}, the JDBC connection string starts with \texttt{"jdbc"} prefix followed by the scheme type, server address, and database name. We observed that the username and password can be present in two forms, either before the server address or separately in the query parameters. We combined the two forms and formulated a common regex for Group 3.

To separate the secret and asset from the database connection string, we used the capturing group~\cite{capturinggroup} feature of regex. The capturing group allows us to capture a specific part of the match. For example, as shown in Table~\ref{regex-database}, we implemented three capturing groups in the MySQL regex. The capturing group \texttt{<dbms>} captures the database type, \texttt{<credentials>} captures the username and password, and \texttt{server} captures the server address of the database.

\textbf{Step 1.2 Identifying Secret-Asset Pairs Using Regex:} In this step, we executed the regexes formulated in Step 1.1 to identify the database connection strings. We used the \texttt{re}~\cite{re} library of Python to execute the regexes. Since the database connection strings can be present in the Git commit history of a repository, we used \texttt{GitPython}~\cite{gitpython}, a Python library for traversing the commit history. In addition to the commit ID, file path, and line number of a match, we extracted the secret and the corresponding database asset from the connection string using the capturing group of regex.

\textbf{Step 2: Asset Finding Using Data Flow Analysis:}\label{DataFlowAnalysis} Among the four patterns described in Section~\ref{AssetPatterns}, except Pattern 1, we observed that the secret and the corresponding database asset are not present in the same string. Instead, the secret and the corresponding database asset are defined separately and passed into a database driver function defined in the same or separate source file from where the secret and asset are present. For example, as shown in Figure~\ref{fig:pattern3} (Pattern 3), the database username, password, and the server address present in lines 3, 4, and 2, respectively, are passed into \texttt{mysql.connect} driver function defined in line 7.

For AssetHarvester, we utilized Data Flow analysis~\cite{khedker2017data} to detect the flow of secrets and assets into the database driver functions. Previous research~\cite{akondresourceprop} has used Data Flow analysis for security weakness propagation in the source code, such as the use of weak cryptographic algorithms. In a Data Flow analysis~\cite{khedker2017data}, the data flow among program elements of the entire source code is modeled through a Data Flow Graph (DFG). A DFG is a directed graph that consists of a set of nodes and a set of edges. The nodes in the DFG represent the semantic elements that carry values at runtime, whereas edges represent the way data flows between program elements. In a program, a node representing the origin of data is called the \texttt{Source}, whereas a node representing the destination of the data is called the \texttt{Sink}. In our study, a database secret and the corresponding asset are the Sources, and the database driver functions are the Sinks. We now describe the process of identifying the Python database drivers for our study. Additionally, we discuss the ways sources can flow into the database driver sinks and the process of identifying the secret-asset pairs from the sources and sinks.

\textbf{Step 2.1 Identifying Database Drivers:} To identify the Python database drivers, we constructed a set of search strings: \emph{(MySQL OR PostgreSQL OR MongoDB OR SQL Server) AND (driver for Python)}. We selected the top 100 results from Google Search Engine for each search string. The stopping criteria for choosing the top 100 results are based on the grey literature search guideline in prior studies~\cite{GAROUSI2019101}. From the search result, we identified 12 database drivers grouped in 7 categories, which are presented in Table~\ref{database-driver-types}. We observed that in addition to identifying database drivers for the four databases, ODBC and JDBC, we identified two drivers, peewee~\cite{peewee} and SQLAlchemy~\cite{sqlalchemy} for the Object Relational Mapper (ORM) framework~\cite{ORM}. ORM is different than other drivers since ORM abstracts the database access with objects instead of directly managing the database access with SQL queries. The identified drivers have a function such as \texttt{connect} or \texttt{create\_pool} to connect with the database. We observe that a driver function can have two different argument types (Positional and Keyword)~\cite{positionalarg}, which act as the sinks for database username, password, and server address. The columns ``Positional Argument" and ``Keyword Argument" of Table~\ref{database-driver-types} indicate which argument type each driver supports. We now discuss the two argument types as sinks for database secrets and assets.

\begin{table}[]
\begin{center}
\caption{List of Python database drivers with their supported arguments for secret-asset pairs}
\label{database-driver-types}
\small
\begin{tabular}{|l|l|c|c|}
\hline
\textbf{Category}           & \textbf{Driver Name}     & \begin{tabular}[c]{@{}c@{}}\textbf{Positional}\\ \textbf{Argument}\end{tabular} & \begin{tabular}[c]{@{}c@{}}\textbf{Keyword}\\ \textbf{Argument}\end{tabular} \\ \hline \hline
\multirow{3}{*}{MySQL}      & aiomysql~\cite{aiomysql}        &              & \checkmark                    \\ \cline{2-4} 
                            & mysql-connector~\cite{mysql-connector} &              & \checkmark                    \\ \cline{2-4} 
                            & PyMySQL~\cite{pymysql}         &  \checkmark             & \checkmark        \\ \hline
\multirow{3}{*}{PostgreSQL} & aiopg~\cite{aiopg}           &   \checkmark            & \checkmark           \\ \cline{2-4} 
                            & asyncpg~\cite{asyncpg}         &   \checkmark            &  \checkmark      \\ \cline{2-4} 
                            & psycopg2~\cite{psycopg2}        &    \checkmark           &   \checkmark      \\ \hline
MongoDB                     & pymongo~\cite{pymongo}         &             &   \checkmark       \\ \hline
SQL Server                  & pymssql~\cite{pymssql}         &              & \checkmark                    \\ \hline
ODBC                        & pyodbc~\cite{pyodbc}          & \checkmark                        &                    \\ \hline
JDBC                        & JayDeBeApi~\cite{jaydeapi}      &  \checkmark                       &                     \\ \hline
\multirow{2}{*}{ORM}        & peewee~\cite{peewee}          & \checkmark           &  \checkmark                  \\ \cline{2-4} 
                            & SQLAlchemy~\cite{sqlalchemy}      &                         &    \checkmark    \\ \hline
\end{tabular}
\end{center}
\end{table}

\begin{figure}
    \includegraphics[width=\columnwidth, frame]{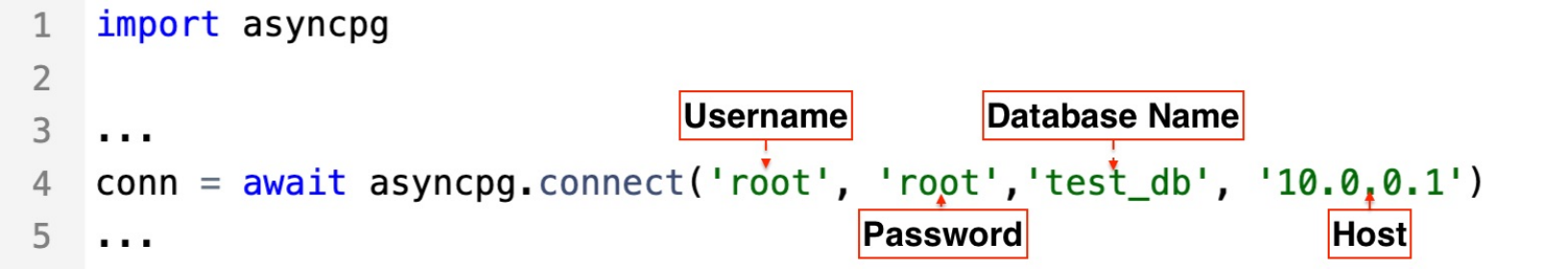}
    \caption{The database credentials and server address are passed in a specific order in the database driver function.}
    \label{fig:positional-argument}  
\end{figure}

\textit{\uline{1. Positional Argument:}} A positional argument~\cite{positionalarg} is passed to a function based on the position in the argument list without explicitly specifying the parameter name. Since the order of the position of the arguments is fixed, we know which positions will act as the database credentials (username and password) and asset (host, port, and database name) sinks. For example, as shown in Figure~\ref{fig:positional-argument}, the username, password, database name, and host address of the database are passed in a specific order in the connect function of \texttt{asyncpg}. Hence, we identified the sources that flow into each ordered position of the driver function for the database secrets and assets.

\begin{figure}
    \centering
    \begin{subfigure}[b]{\columnwidth}
        \centering
        \includegraphics[width=\columnwidth, frame]{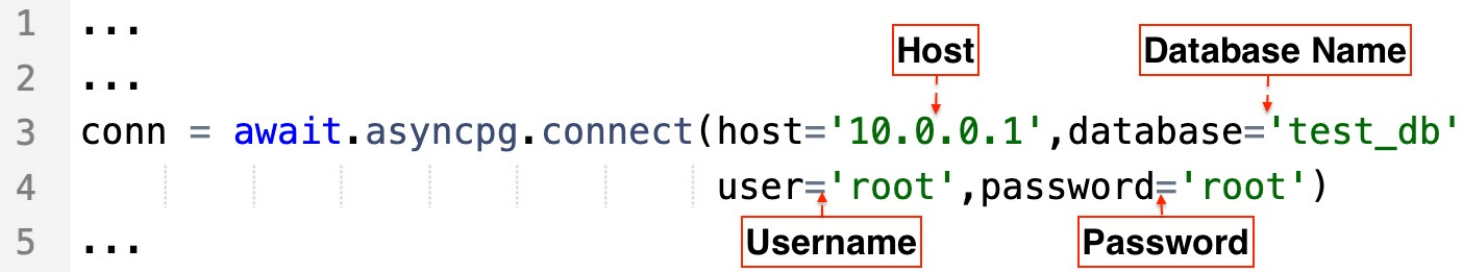}
        \caption[]{{Keyword arguments passed as separate parameters}}
        \label{fig:key-arg-1}
    \end{subfigure}
\vskip\baselineskip
    \begin{subfigure}[b]{\columnwidth}
        \centering
        \includegraphics[width=\columnwidth, frame]{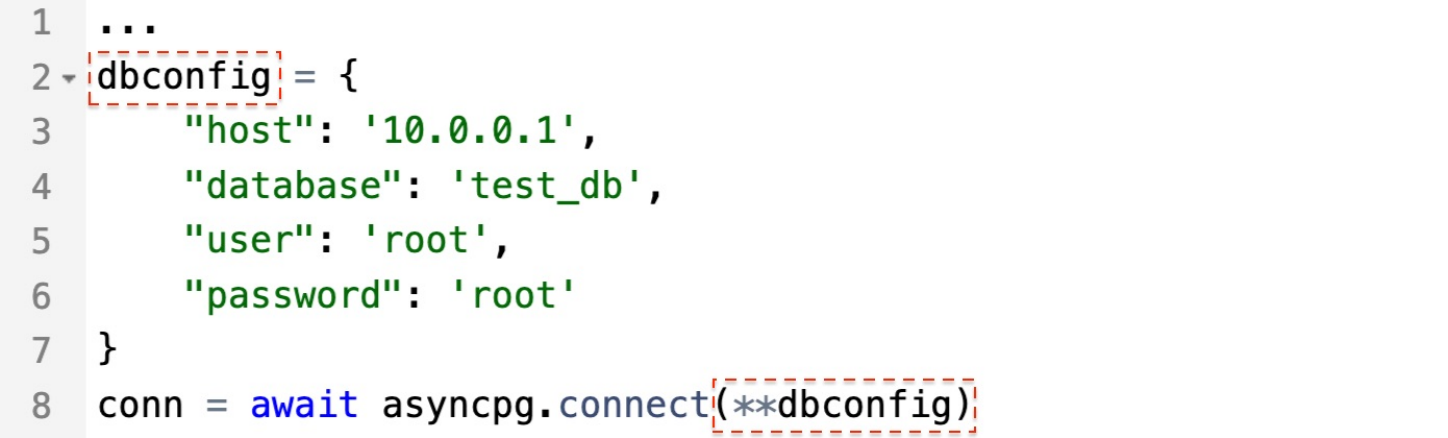}
        \caption[]{{Keyword arguments defined in a dictionary}}
        \label{fig:key-arg-2}
    \end{subfigure}
\caption{The database credentials and server address are passed as keyword arguments in the database driver functions.}
\label{fig:key-args}
\end{figure}

\textit{\uline{2. Keyword Argument:}} A keyword argument~\cite{positionalarg} (also called Named argument) is passed to a function by specifying the parameter name with the corresponding value. Unlike positional argument, the order of keyword argument is not fixed in a function. We observe that keyword arguments can be passed in separate parameter names and dictionary objects. As shown in Figure~\ref{fig:key-arg-1}, the username, password, database name, and host address of the database are passed in separate named arguments without fixed order, whereas defined in a dictionary object and passed in the function as shown in Figure~\ref{fig:key-arg-2}. Since we know the argument names, we can identify the sources flowing into the relevant arguments of the driver function for the database secrets and assets.

\textbf{Step 2.2 Identifying Secret-Asset Pairs Using CodeQL:} 
For Data Flow analysis, we used Version \texttt{2.15.1} of CodeQL~\cite{codeql}, an open-source source code analysis framework developed by GitHub. CodeQL treats source code as data and creates databases containing a hierarchical representation of the code, such as the abstract syntax tree, the data flow graph, and the control flow graph. Developers can query the database using QL~\cite{QL}, a query language optimized for efficiently analyzing databases representing software artifacts~\cite{QL}. First, we queried the abstract syntax tree to identify the database driver sinks (functions and respective arguments). Next, we queried the data flow graph to find the sources that fall into the identified sinks and find the value of the secret-asset pairs for the corresponding sources using the abstract syntax tree. Finally, we queried the control flow graph to find the location (file path and line number) of secret-asset pairs. Since the database drivers are external libraries, we utilized the API Graphs~\cite{apigraph} of CodeQL to compute the data flow graph. API Graphs are a uniform interface for referring to functions, classes, and methods defined in external libraries. In our study, we used the \texttt{semmle.python.ApiGraphs} module for accessing the external library functions.

\begin{table*}[!ht]
\begin{center}
\caption{Statistics of the presence of database assets in the neighboring lines of the secrets of the same file in AssetBench}
\label{database-secret-asset-location}
\small
\begin{tabular}{|c|ccccc|}
\hline
\multirow{2}{*}{\textbf{Secret}} & \multicolumn{5}{c|}{\textbf{\begin{tabular}[c]{@{}c@{}}Absolute Difference Between Secret and Asset Line Number \\ (Number of Secret-Asset Pairs)\end{tabular}}} \\ \cline{2-6} 
 &
  \multicolumn{1}{c|}{\textbf{0}} &
  \multicolumn{1}{c|}{\textbf{1}} &
  \multicolumn{1}{c|}{\textbf{2}} &
  \multicolumn{1}{c|}{\textbf{3}} &
  \textbf{\textgreater{}=4} \\ \hline
Database Password &
  \multicolumn{1}{c|}{407 (31.9\%)} &
  \multicolumn{1}{c|}{340 (26.7\%)} &
  \multicolumn{1}{c|}{349 (27.4\%)} &
  \multicolumn{1}{c|}{124 (9.7\%)} &
  54 (4.2\%) \\ \hline
\end{tabular}
\end{center}
\end{table*}

\begin{figure}
    \includegraphics[width=\columnwidth, frame]{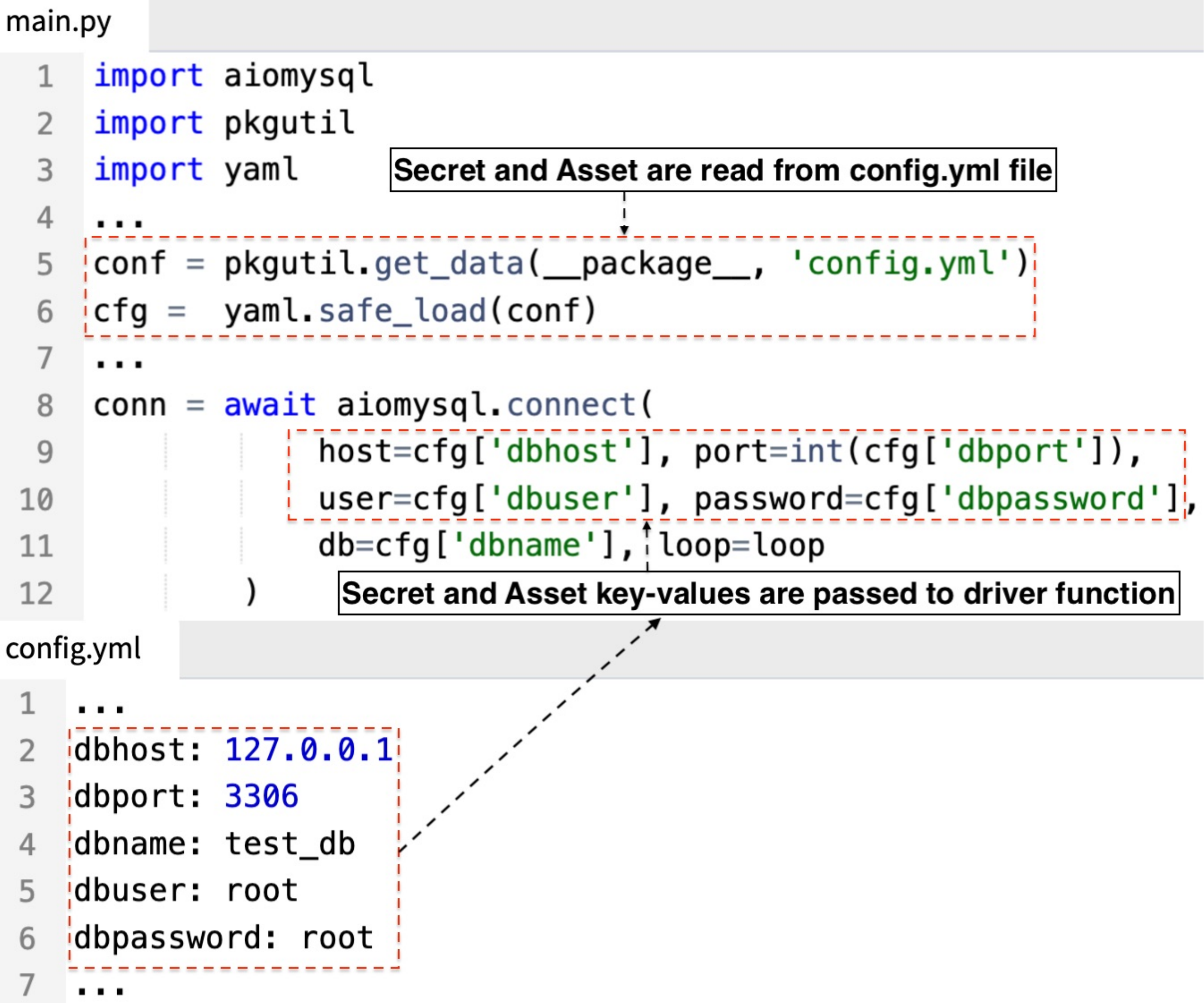}
    \caption{The config.yml file contains the database secret-asset pair that is read in the main.py file. The secret-asset values are accessed by the key names and passed to the driver function.}
    \label{fig:asset-config-file}  
\end{figure}

\textbf{Step 2.3 Identifying Secret-Asset Pairs Using CodeQL and File Parsing:}\label{fileparsing} We observed that the database secret and the corresponding asset can be present in a configuration (config) file such as YAML, JSON, and XML files. The config file is read as a dictionary object, and the values of the dictionary object are accessed in the driver function. For example, as shown in Figure~\ref{fig:asset-config-file}, the secret and the corresponding asset of MySQL database are present in the \texttt{config.yml} file, which is read in a dictionary object \texttt{cfg} of the \texttt{main.py} file (lines 5 and 6). The values of dictionary object \texttt{cfg} are accessed in the \texttt{aiomysql.connect} driver function (lines 8-11) using key names such as \texttt{dbhost} and \texttt{dbuser}. However, CodeQL does not support data flow analysis of source codes across multiple programming languages. As a result, the flow of secrets and assets from the \texttt{config.yml} file into the driver function of the \texttt{main.py} file can not be captured. 

However, we observed that by utilizing the data flow analysis of CodeQL, we can find the config file name and the key names that flow into the driver function. Since we identified the config file name and associated the key names, we parsed the config file and retrieved the values for each key name. For retrieving the values from the YAML, JSON, and XML files, we used the \texttt{PyYAML}~\cite{PyYAML}, \texttt{json}~\cite{json} and \texttt{xmltodict}~\cite{xmltodict} packages of Python, respectively. 

\textbf{Step 3: Asset Finding Using Fast-Approximation Heuristic:} We observed that developers may have accidentally or intentionally kept the secret and the corresponding asset as commented lines in the source code. However, commented lines are ignored during data flow analysis. Additionally, capturing the data flow may not always be possible if the source code has dynamic behavior, such as extensive use of reflection. Thus, we can not identify the assets protected by secrets in those cases using data flow analysis. However, when the secret-asset pair is present in the same file, we observed from AssetBench that the database asset may be present in the neighbor lines of the corresponding secret. As shown in Table~\ref{database-secret-asset-location}, the percentage of database assets present within three neighboring lines of the corresponding database password in the same file is 95.8\%. Thus, we can check the neighboring lines of the secret line to identify the corresponding asset. In our study, we define three neighboring lines as three lines above and three lines below the secret line. For example, if a secret is present in line 20, the asset can be present between line 17 and line 23. We now discuss the approach of identifying the secret-asset pairs using neighboring lines.

\textbf{Step 3.1 Identifying and Filtering Secrets:} First, we identified the secrets in the repositories using two open-source secret detection tools, TruffleHog~\cite{trufflehog} and Gitleaks~\cite{gitleaks}. The authors of SecretBench~\cite{secretbench} have used the same two tools to curate the benchmark dataset as discussed in Section~\ref{BenchmarkDataset}. Since the two tools can overlap outputting the same database secret in a repository, we merged the unique secrets. Next, we filtered the unique secrets for which we already found assets using Regex (Step 1) and Data Flow Analysis (Step 2).

\begin{figure}
    \includegraphics[width=\columnwidth, frame]{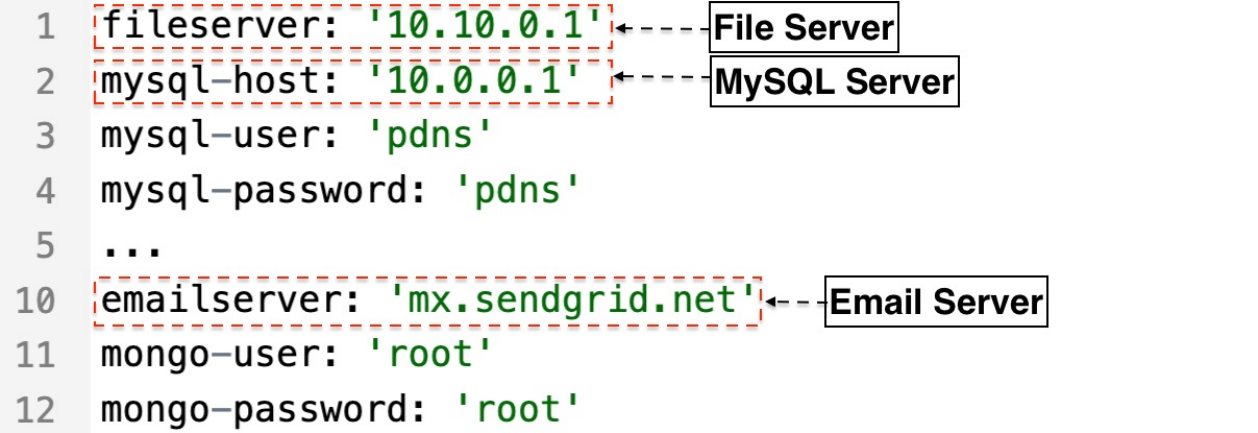}
    \caption{Multiple or zero corresponding assets can be present in the neighboring lines of a secret.}
    \label{fig:neighboring-lines}  
\end{figure}

\textbf{Step 3.2 Identifying Secret-Asset Pairs Using Neighboring Lines:} In this step, to identify the neighboring lines for each secret, we used the \texttt{linecache}~\cite{linecache} library of Python that provides random access to source code lines. We observe that a database asset identifier can be present as an IP address (\texttt{\seqsplit{10.0.0.1}}) or a DNS name (\texttt{\seqsplit{wrpxdb.bioch.edu}}), as shown in Pattern 2 and 4, respectively. Hence, we formulated regexes for capturing the IP addresses (\verb@\b(?:\d{1,3}\.){3}\d{1,3}\b@) and DNS names (\verb@\b[A-Za-z0-9][A-Za-z0-9-.]*\.\D{2,4}\b@) in the neighboring lines. However, the neighboring lines may contain multiple IP addresses and DNS names. Among those assets, one asset can be the corresponding asset for a secret. For example, as shown in Figure~\ref{fig:neighboring-lines}, a file server and a MySQL database address are present in lines 1 and 2, respectively. The correct asset for the MySQL database username and password present in lines 3 and 4 is the MySQL database address. In addition, the asset protected by the secret may not be present in the source code. For example, a DNS name for an email server is present in line 10. However, the email server is not the asset protected by the MongoDB database username and password present in lines 11 and 12, respectively.

We observe that a specific group's secret and corresponding asset can have the same prefix in the variable or key names. For example, as shown in Figure~\ref{fig:neighboring-lines}, the key names of username (\texttt{mysql-user}), password (\texttt{mysql-password}) and server address (\texttt{mysql-host}) of MySQL database have the same prefix (\texttt{mysql}). However, the key name of the file server does not start with the same prefix as the key names of the MySQL database. Hence, we can apply a string-matching algorithm to calculate similarity scores between the secret line and the candidate asset lines and choose the asset with the highest similarity score. In addition, we discard the asset if the similarity score with the secret line is less than a threshold. To calculate the similarity score, we used Jaro-Winkler Similarity~\cite{winkler1990string}, a string-matching algorithm that uses a prefix scale by giving a high similarity score to strings that match from the beginning. The Jaro–Winkler algorithm provides a similarity score between 0 and 1, and we chose 0.5 as the threshold similarity score. We utilized the \texttt{jaro\_winkler\_similarity} function from the \texttt{jellyfish}~\cite{jellyfish-python} package in Python to compute the similarity score and identify the secret-asset pairs.

\section{Performance of AssetHarvester} \label{Results}
In this section, we answer RQ2 by evaluating the performance of AssetHarvester against AssetBench.

\textbf{Precision, Recall and F1-Score:} Table~\ref{precision-recall} presents the precision, recall and F1-score of AssetHarvester for each database type. The column ``Precision (TP, FP)" denotes the precision for each database type. The number in parenthesis denotes the number of true positive and false positive secret-asset pairs outputted by AssetHarvester. The column ``Recall (TP, FN)" denotes the recall for each database type. The number in parenthesis denotes the number of true positive and false negative secret-asset pairs outputted by AssetHarvester. The column ``F1 Score" denotes each database type's F1-score (the harmonic mean of precision and recall). We now discuss our observations related to precision, recall, and F1-score.

\begin{table}[]
\begin{center}
\caption{Precision, Recall and F1-score of AssetHarvester for each database type}
\label{precision-recall}
\small
\begin{tabular}{|l|l|l|l|}
\hline
\textbf{Database Type} &
  \multicolumn{1}{c|}{\textbf{\begin{tabular}[c]{@{}c@{}}Precision \\ (TP, FP)\end{tabular}}} &
  \multicolumn{1}{c|}{\textbf{\begin{tabular}[c]{@{}c@{}}Recall \\ (TP, FN)\end{tabular}}} &
  \multicolumn{1}{c|}{\textbf{\begin{tabular}[c]{@{}c@{}}F1\\ Score\end{tabular}}} \\ \hline \hline
\textbf{MySQL}      & 0.98 (712, 13)  & 0.91 (712, 65)   & 0.94 \\ \hline
\textbf{PostgreSQL} & 0.98 (620, 10)  & 0.91 (620, 51)   & 0.94 \\ \hline
\textbf{MongoDB}    & 0.96 (286, 11)  & 0.92 (286, 24)   & 0.94 \\ \hline
\textbf{SQL Server} & 1.00 (8, 0)     & 0.32 (8, 17)     & 0.48 \\ \hline
\textbf{Overall}    & \cellcolor[HTML]{C0C0C0}0.97 (1626, 34) & \cellcolor[HTML]{C0C0C0}0.90 (1626, 165) & \cellcolor[HTML]{C0C0C0}0.94 \\ \hline
\end{tabular}
\end{center}
\end{table}

\begin{itemize}
    \item We observed that AssetHarvester demonstrated overall 97\% precision, indicating high precise detection of assets protected by secrets with low false positives. The count of false positives (34) indicates that the tool incorrectly outputted 34 assets out of 1,791 secret-asset pairs.
    \item The overall recall score of AssetHarvester is 90\%, indicating a strong ability to identify instances of assets for secrets, supported by an F1-score of 94\%. The count of false negatives (165) indicates that the tool failed to detect 165 instances of secret-asset pairs.
    \item Among the four database types, the recall score of SQL Server is low (32\%) though the precision score is 100\%. The tool could not detect 17 instances of SQL Server assets out of 25 secret-asset pairs. We discussed the reason for false negatives later in this section.
\end{itemize}

\begin{figure}
    \includegraphics[width=\columnwidth]{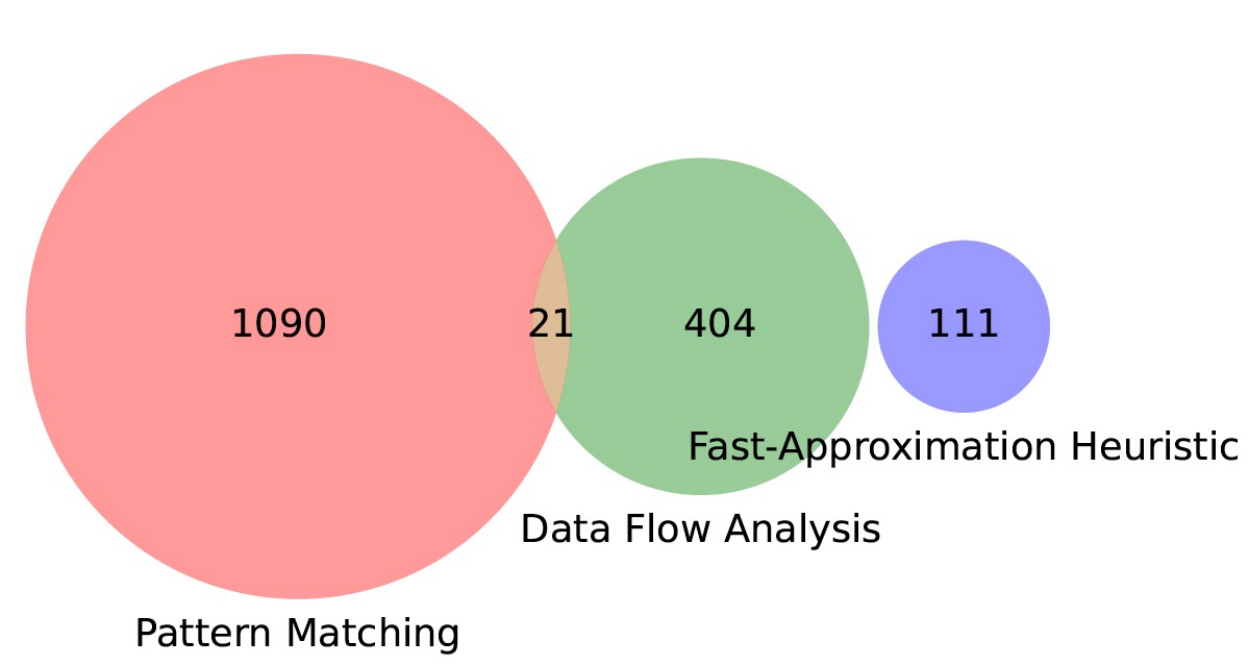}
    \caption{The number of unique secret-asset pairs found by Pattern Matching, Data Flow Analysis, and Fast-Approximation Heuristic approaches.}
    \label{fig:approach-performance}  
\end{figure}

\textbf{Performance of Pattern Matching, Data Flow Analysis, and Fast-Approximation Heuristic:}
Figure~\ref{fig:approach-performance} depicts that AssetHarvester detected unique secret-asset pairs using the three approaches, thus indicating the importance of the three approaches. Out of 1,626 secret-asset pairs, using pattern matching (regex) and data flow analysis (CodeQL), we found 1,090 and 404 unique secret-asset pairs, respectively. In addition, we found 111 unique secret-asset pairs using the fast-approximation heuristic (Neighboring Lines). However, we observed that 21 instances of secret-asset pairs were detected by both pattern matching and data flow analysis. The overlap happened because of \texttt{dsn} keyword argument of driver functions, which takes a connection string that is also matched by the regex of database types. However, the overlap is low since all the connection strings found by the regex are not passed to Python database driver functions. Instead, the connection strings are either passed to non-python such as Java or .NET database driver functions or not passed to any functions. Thus, those connection strings could not be captured by data flow analysis. Additionally, we observed that among the three approaches, AssetHarvester incorrectly detected 9 and 25 secret-asset pairs out of 34 false positives using pattern matching and fast-approximation heuristics, respectively. However, AssetHarvester did not detect any false positives using data flow analysis since we used specific sinks for database secret-asset pairs from the documentation compared to generic sinks implemented by GitHub CodeQL~\cite{codeql-generic-sink}. Additionally, we filtered sources flowing into sinks that do not point to primitive values (string or integer). We also filtered values with only colons or slashes that flowed into the sinks as part of the asset URL from string concatenation to avoid false positives. We now discuss our observations on the rules triggering the false positives and false negatives.

\uline{Analysis of False Positives:} We observed that the false positives are mostly triggered by the neighboring lines rule (73.5\% of the 34 false positives reported by AssetHarvester). We noticed that the key names of the secret and corresponding asset do not always follow the specific pattern of having similar prefixes (Step 3.2, Section~\ref{AssetHarvester}). For example, ``URL" and ``password" are the key names of a database server address and password but do not have the same prefixes. As a result, when multiple IP addresses or DNS names were present in the neighboring lines, and the similarity score met the threshold, AssetHarvester could not detect the correct asset for a secret.

\uline{Analysis of False Negatives:} We observed that AssetHarvester failed to detect secret-asset pairs when the asset is not present within three neighboring lines of the secret. As shown in Table~\ref{database-secret-asset-location}, 54 (4.2\%) instances of secret-asset pairs in AssetBench do not fall within three neighboring lines. In our study, the repositories also contain non-Python source codes, such as Java and .NET, where the secret-asset pairs are passed to Java and .NET database driver functions. However, AssetHarvester did not detect those secret-asset pairs since we only executed data flow analysis for Python source codes in our study. Thus, AssetHarvester shows a lower recall (32\%) for SQL Server among other database types since the SQL Server assets are typically passed to .NET driver functions. Additionally, AssetHarvester could not detect assets present as variables in the connection strings not passed to Python driver functions. For example, the connection string \texttt{\seqsplit{"jdbc:postgresql://\$\{databaseServer\}"}} contains the variable \texttt{databaseServer} defined separately with the actual value.

\textbf{Comparison with Baseline Secret Detection Tools:} Existing secret detection tools do not detect the assets protected by secrets. Thus, we could only compare AssetHarvester's performance on secret detection with the existing tools. Basak et al.~\cite{basak2023compare} compared five open-source and four proprietary secret detection tools against SecretBench. We selected these nine baseline tools and evaluated them against 188 repositories containing 2,114 secrets of four databases, as curated in Section~\ref{BenchmarkDataset}. Table~\ref{baseline-comparison} presents each tool's precision and recall. We observed that the nine baseline tools show lower precision (less than 50\%) and recall (less than 41\%) scores than AssetHarvester (precision 92\% and recall 90\%). Additionally, we noticed that GitHub-scanner did not output any database secrets since the supported secret patterns lack database secret patterns~\cite{github-secret-patterns}. However, GitHub introduced generic secret scanning for unstructured secrets such as database passwords, which we could not compare since the tool is in the beta phase and restricted to enterprise accounts only~\cite{github-generic-secret}.

We observed that the lower precision of the tools is due to employing generic regex. For example, TruffleHog detects the database connection string but does not check if the connection string contains a password, thus outputting the connection string without a password as a secret. We resolved these false positives using the capturing group of regex for AssetHarvester (Step 1.2, Section~\ref{AssetHarvester}). Additionally, the lower recall is due to employing insufficient rulesets and ineffective entropy calculation. For example, tools reject a secret based on entropy score. However, we found instances of secrets having lower entropy scores protecting real assets. We improved recall by identifying secrets flowing into the respective database driver functions using data flow analysis without considering the entropy score for AssetHarvester (Step 2, Section~\ref{AssetHarvester}). In addition, we identified 86 secrets that are not present in SecretBench using data flow analysis. As discussed in Section~\ref{BenchmarkDataset}, the authors of SecretBench used two open-source tools, TruffleHog and Gitleaks, to curate the benchmark dataset. These tools leverage regex and entropy scores to identify secrets. Thus, secrets that are not matched by the regex and entropy scores are missed.

\begin{table}[]
\begin{center}
\caption{Comparison of AssetHarvester with 9 Baseline Secret Detection Tools on Secret Detection}
\label{baseline-comparison}
\small
\begin{tabular}{|l|l|l|l|}
\hline
\textbf{Tool} &
  \multicolumn{1}{c|}{\textbf{\begin{tabular}[c]{@{}c@{}}Precision \\ (TP, FP)\end{tabular}}} &
  \multicolumn{1}{c|}{\textbf{\begin{tabular}[c]{@{}c@{}}Recall \\ (TP, FN)\end{tabular}}}\\ \hline \hline
git-secrets~\cite{git-secret}  & 0.04 (1460, 65)  & 0.01 (31, 2083) \\ \hline
Gitleaks~\cite{gitleaks}      & 0.21 (220, 45)  & 0.02 (37, 2077) \\ \hline
Repo-supervisor~\cite{repo-supervisor}    &  0.31 (1270, 391)  & 0.17 (364, 1750) \\ \hline
TruffleHog~\cite{trufflehog} & 0.19 (5666, 1068)     & 0.40 (851, 1263) \\ \hline
Whispers~\cite{whispers} & 0.06 (2203, 147)     & 0.05 (112, 2002)  \\ \hline
Commercial X & 0.35 (7260, 2511)    & 0.28 (589, 1525)  \\ \hline
ggshield~\cite{ggshield} & 0.49 (1972, 969)    & 0.16 (330, 1784)  \\ \hline
GitHub-scanner~\cite{github-secret-scanner} & 0.00 (0, 0)    & 0.00 (0, 0)     \\ \hline
Spectralops~\cite{spectralops} & 0.15 (574, 88)    & 0.29 (609, 1505)    \\ \hline
\cellcolor[HTML]{C0C0C0}AssetHarvester    & \cellcolor[HTML]{C0C0C0}0.92 (2298, 89) & \cellcolor[HTML]{C0C0C0}0.90 (1892, 222)\\ \hline
\end{tabular}
\end{center}
\end{table}

\textbf{Developer Survey:} Since we leveraged a random sample of the dataset to construct AssetHarvester, AssetHarvester's evaluation against AssetBench is susceptible to bias. However, no other publicly-available benchmark dataset containing secret-asset pairs is present. Basak et al.~\cite{secretbench} initially curated 89,070 candidate GitHub repositories for SecretBench. Since identifying and manually labeling secrets from 89,070 repositories was impractical, they finally selected 818 repositories using a multiset-multicover algorithm~\cite{secretbench}. To mitigate bias, we selected a random sample of 15 repositories from 88,252 candidate repositories (excluding 818 repositories of SecretBench) after applying Criteria 1 and Criteria 2 as discussed in Section~\ref{BenchmarkDataset}. We identified 42 secret-asset pairs (18 unique secret-asset pairs) present in repositories' commit history using AssetHarvester. Then, we conducted a developer survey with the committer of secret-asset pairs to evaluate AssetHarvester's performance, and 18 responses were received. Fifteen respondents agreed with our identified secret-asset pairs, and three respondents termed the secret-asset pairs as false positives. 

\textbf{AssetHarvester's Effectiveness Beyond Benchmark:} We identified the secret-asset co-location patterns from a random sample of AssetBench containing database secret-asset pairs (Section~\ref{AssetPatterns}). However, we selected a random sample of 10 secret-asset pairs for each of 7 non-database secret types, such as API keys, tokens, and private keys from SecretBench~\cite{secretbench}. We found that the co-location of all the selected 70 secret-asset pairs matches our identified four co-location patterns, thus indicating the generality of the identified patterns. For example, as depicted in Pattern 1 (Same String, Same Line, Same File), the secret-asset pair can be present in the same string in a URL. Thus, we can identify the API key and corresponding server endpoint from the API URL, similar to a database connection string. Additionally, we did not limit AssetHarvester to the database drivers found in the random sample. Instead, we identified the database drivers from the database provider documentation. We found instances of only 3 database drivers in the random sample. However, we constructed AssetHarvester with 12 database drivers and detected secret-asset pairs from 9 database drivers in AssetBench. We also identified secret-asset pairs from one new database driver while evaluating AssetHarvester's performance with 15 new repositories (Section~\ref{Results}).

\section{Discussion} \label{Discussion}
In this section, we discuss the implications of AssetHarvester from the findings of our study.



\textbf{Data Flow Analysis aids in detecting all parts of a credential and the corresponding asset as one group.} A credential can have multiple parts required to access the protected asset. For example, an access key ID and a secret access key are required to access an AWS resource, whereas for accessing a database, both a username and a password are required. Similar to credentials, assets can have multiple parts as well. For example, a database asset consists of a host, port, and database name. However, existing secret detection tools can not detect all parts of a credential if present separately in the source code. In addition, the tools output each part of a credential in separate alerts instead of outputting as one group. Thus, developers need to manually identify the related alerts out of all the alerts reported by the tools. However, AssetHarvester leverages data flow analysis to detect all parts of a credential and the asset and output as one alert to the developers. In our study, we detected the database credential (username and password) and asset (host, port, database name) flowing into the same database driver functions using data flow analysis. Additionally, we provided the information on the call location as an additional context for the developers to prioritize secret and asset eradication from the source code. 

\textbf{AssetHarvester can be extended to detect secret-asset pairs in other programming languages and non-database secret types.} In our study, we detected secret-asset pairs of four database providers in Python programming language. However, the found secret-asset co-location patterns are generic and can be applied to other programming languages and non-database secret-asset pairs (Section~\ref{Results}). We now discuss the effort needed and challenges to extend AssetHarvester for other programming languages and secret types.

\textit{\uline{Programming Languages:}} Among the three techniques (pattern matching, data flow analysis, and fast approximation heuristics) we leveraged, pattern matching and fast approximation heuristics are programming language-agnostic. Hence, these two techniques can be applied to find secret-asset pairs in other programming languages without additional effort. On the contrary, the data flow analysis is dependent on the programming language. However, the abstract syntax tree, control flow, and data flow graph of the source code for each language in a repository can be computed separately, which CodeQL supports. Then, the queries to identify secret-asset pairs can be run on each of the computed graphs, thus extending for each programming language with minimal effort.

\textit{\uline{Non-Database Secret Types:}} We observed a random sample of 10 secrets for seven secret types such as private key, API key, and authentication token from SecretBench~\cite{secretbench}. Next, we categorized the secret-asset pairs into two categories. We now describe how we can extend AssetHarvester to detect secret-asset pairs except for the database provider.

\textbf{1. Cloud Providers:} For authentication and authorization with cloud providers, such as Google Cloud, API keys and tokens are used that can be present in an API URL or separately passed to a function. Since each cloud provider follows a specific format for API URLs, a regex can be formulated for each API and added to the regex list (Step 1.1, Section~\ref{PatternMatching}). Additionally, when the secret is not present in the API URL, the secret and the API URL can be detected by data flow analysis since these values are passed in a common HTTP request client (\texttt{get} and \texttt{post} methods). We observed that cloud secrets are the second most exposed secrets on GitHub, according to the 2024 GitGuardian report~\cite{gitguardian-secret-sprawl}. Additionally, they found a 1212-fold increase in leaked OpenAI API keys since 2022, driven by the rising use of large language models (LLMs). We will prioritize extending AssetHarvester to identify cloud provider secrets for future work.   

\textbf{2. Non-Database Servers:} A secret can protect non-database servers, such as Mail and FTP servers. Similar to database servers (Group 1, Table~\ref{regex-database}), non-database servers have specific formats containing a scheme type (\texttt{\seqsplit{scheme://user:password@host:port}}). For example, the ``smtp" or ``pop3" are the Mail server's scheme types, whereas ``ftp" is the scheme type for the FTP server. We can add the scheme types in Group 1 of the regex list to capture the non-database server secret-asset pairs. In addition, the secret and the corresponding server URL are used by functions such as ``login" and ``SMTP" functions of \texttt{smtplib}~\cite{smtplib} module of Python for sending email. Web and Mail servers also use private keys to enable secure connections. These keys are stored in a file separately from the asset location and read from another function in the source code. Thus, we can leverage data flow analysis to retrieve the file name from the function and parse the file to identify the private key, as shown in Step 2.3 of Section~\ref{AssetHarvester}.

Our list of regexes and sinks for AssetHarvester is configurable and requires no source code change to detect non-database secret types. Though identifying the regex and sinks for each secret type from the documentation requires manual effort, the process can be automated in the future. For example, LLMs can aid in identifying regex and sinks from source code patterns and vendor documentation for each secret type. We can leverage the knowledge gained from the manual analysis of our study and generate prompts for LLMs.

\section{Ethics and Data Protection} \label{Ethics}
Since our dataset contains sensitive information, we will distribute the dataset selectively. Researchers and tool developers who want to use our dataset will sign a data protection agreement with us to ensure ethical use. In addition, we did not attempt to use the secret-asset pairs to verify their validity. We only contacted the developers who committed the secret-asset pairs to validate our labeling. Additionally, we are notifying every developer in our dataset to remove the secret-asset pairs from their VCS. 

\section{Threats to Validity} \label{ThreatToValidity}
In this section, we discuss the limitations of our paper. 

\textit{\uline{Manual Analysis}}: Manual analysis can introduce bias due to the multiple interpretations and oversights. For example, identifying the assets protected by secrets while curating AssetBench is susceptible to bias. We mitigated the bias by cross-checking the identified secret-asset pairs with two raters.

\textit{\uline{Benchmark Dataset}}: Our selection of benchmark dataset for secrets is susceptible to bias. Basak et al.~\cite{secretbench} utilized two open-source tools (TruffleHog and Gitleaks) and curated SecretBench, which we extended as AssetBench by identifying the protected asset for each secret. However, we observed that AssetHarvester identified 86 secrets not present in SecretBench (Section~\ref{Results}). Thus, SecretBench may have more missing secrets, impacting the results discussed in Section~\ref{Results}.

\textit{\uline{Developer Survey}}: For the developer survey of AssetBench, we selected the secret-asset pairs committed between 2021 and 2022. However, the developer's responses could have recall bias. To mitigate the bias, we provided screenshots of the secret and asset-containing source code with metadata (commit ID, file path, and line number) to the developers.

\textit{\uline{Data Flow Analysis}}: In our study, we used CodeQL for data flow analysis in the latest snapshot of the repositories. CodeQL can only model the data flow with the provided snapshot of the source code. However, developers can push secret-asset pairs in one commit and remove them in another commit. Secret-asset pairs can still be present in the old snapshot, that can not be detected by executing data flow analysis on the latest snapshot. However, we can compute data flow analysis for each repository snapshot to identify the secret-asset pairs from Git history, which will be impractical and time-consuming.

\textit{\uline{Neighboring Lines:}} Our selection of three neighboring lines for identifying the assets for a corresponding secret poses a threat to internal validity since the three-line range is selected from AssetBench containing only database secret-asset pairs. However, we selected a random sample of 50 non-database secrets from SecretBench and identified the corresponding assets. We found that the percentage of assets present within three neighboring lines of the corresponding secret is 96.3\%, thus indicating the rule's generalizability to other secret types.

\section{Related Work} \label{RelatedWork}
Previous studies~\cite{meli2019bad,rahman2019share,rahman2019seven, igibeksecret, rahman2021different, krause2023pushed} have investigated the underlying causes of the exposure of secrets in software artifacts. Researchers have found that keeping hard-coded secrets in software artifacts is the most prevalent insecure practice that developers adopt, causing secret leakage. In 2019, Meli et al.~\cite{meli2019bad} found over 100K hard-coded secrets by studying a 13\% snapshot of public GitHub repositories. Rahman et al.~\cite{rahman2019seven} investigated 5,232 Infrastructure as Code (IaC) scripts extracted from 293 open-source repositories and observed 7 ``Security Smells". Among the 7 security smells, hard-coded secrets were found to be the most frequent, with 1,326 occurrences. Within GitHub Gists, which developers use for sharing code snippets, Rayhanur et al.~\cite{rahman2019share} found 689 instances of hard-coded secrets by investigating 5,822 Python Gists in GitHub. These previous works indicate that hard-coded secrets have been leaking in various forms within software artifacts.

Researchers~\cite{krause2023pushed,basaksecretpractice, basakchallenges} have recommended that developers follow secure practices for secret management to avoid exposure of secrets in software artifacts. In 2022, Basak et al.~\cite{basaksecretpractice} identified 24 developer and organization practices by conducting a grey literature review of Internet artifacts such as blog posts. To avoid the accidental commit of secrets, they suggested using VCS scan tools. In another study, Basak et al.~\cite{basakchallenges} investigated the challenges developers face for checked-in secrets in Stack Exchange (SE) and the solutions SE users suggest to mitigate the challenge. They found that to avoid accidentally committing secrets, SE users also suggested using VCS scan tools. However, Basak et al.~\cite{basak2023compare} compared 5 open-source and 4 proprietary VCS scan tools against SecretBench~\cite{secretbench} and observed that tools output a lot of false positives. In addition, tools failed to detect all the secrets present in a repository. Recent research~\cite{saha2020secrets, fengsecret, secrethunter, konygin2023using} has employed Machine Learning (ML) algorithms to reduce the false positives. However, among the 9 VCS scan tools investigated by Basak et al.~\cite{basak2023compare}, two tools (Commercial X (anonymized) and SpectralOps~\cite{spectralops}) employed ML algorithms to detect secrets showed lower precision scores of 25\% and 1\%, respectively. Rayhanur et al.~\cite{rahman2022secret} conducted a developer survey in an XTech company (anonymized) and found that developers ignore secrets due to many secrets outputted by VCS scan tools and time pressure. However, if the information about the asset protected by the secret was provided, developers could have prioritized secret eradication. However, existing VCS scan tools do not provide the asset information for a secret. In this study, we concentrated our research efforts on identifying the assets protected by secrets to aid developers in prioritizing secrets removal efforts. 

\section{Conclusion} \label{Conclusion}
We constructed AssetHarvester, a static analysis tool to detect the assets protected by the corresponding secrets in a repository by investigating the secret-asset co-location patterns. We utilized pattern matching, data flow analysis, and fast-approximation heuristics to construct AssetHarvester. To evaluate AssetHarvester, we curated AssetBench, a benchmark dataset of 1,791 secret-asset pairs comprising four database types. The secret-asset pairs are extracted from 188 public GitHub repositories. We found that AssetHarvester demonstrates precision of (97\%), recall (90\%), and F1-score (94\%) in detecting secret-asset pairs. Our findings indicate that data flow analysis employed in AssetHarvester detects secret-asset pairs with 0\% false positives and also aids in improving the recall of secret detection tools. In addition, though fast-approximation heuristics introduce relatively more false positives, this approach improves recall by detecting assets that cannot be detected using other approaches.


\section*{Acknowledgment}

This work was supported by the National Science Foundation 2055554 grant.




\bibliographystyle{IEEEtran}
%
\bibliography{bibliography}



\end{document}